\documentclass[a4paper, 10pt]{article}

\usepackage[latin1]{inputenc}
\usepackage[dvips]{graphicx} 
\usepackage{epsfig}
\usepackage{times, mathptmx}
\usepackage[square, sort&compress]{natbib}
\usepackage{setspace}
\usepackage{moon}
\usepackage{caption}
\usepackage{fancyhdr}
\usepackage[english]{babel}
\usepackage{amsmath}

\title{ELVIS -- ELectromagnetic Vector Information Sensor}
\author{O. St{\aa}l$^1$, J. Bergman$^1$, B. Thid\'e$^{1,2}$, L. Åhl\'en$^1$ and G. Ingelman$^3$}
\author{J.~E.~S.~Bergman$^1$, L.~{\AA}hl\'en$^1$, O.~St{\aa}l$^1$, 
B.~Thid\'e$^{1,2}$,\\ S.~Ananthakrishnan$^4$,  J.-E.~Wahlund$^1$,
R.~L.~Karlsson$^1$,\\ W.~Puccio$^1$, T.~D.~Carozzi$^3$, and P.~Kale$^5$}

\affiliation{$^1$ Swedish Institute of Space Physics, Uppsala, Sweden\\
\vspace{6pt}
	     $^2$ LOIS Space Centre, V\"axj\"o University, V\"axj\"o, Sweden\\
\vspace{6pt}	          
             $^3$ Space Science Centre, Sussex University, Brighton, UK\\
\vspace{6pt}	          
             $^4$ National Centre for Radio Astrophysics,\\
\                 Tata Institute of Fundamental Research\\ 
\                 Pune University Campus, Ganeshkhind, Pune, India\\
\vspace{6pt}	          
             $^5$ Integrated Circuit and information Technology Pvt. Ltd,\\
                  Department of Electronic Science,\\
\                 University of Pune, Ganeshkhind, Pune, India\\
		  }

\date{}	
\begin{document}
\thispagestyle{fancyplain}
\maketitle
\begin{abstract}
The ELVIS instrument was recently proposed by the authors for the Indian 
Chandra\-yaan-1 mission to the Moon and is presently under consideration
by the Indian Space Research Organisation (ISRO). The scientific objective
of ELVIS is to explore the electromagnetic environment of the moon. 
ELVIS 
samples the full three-dimensional (3D) electric field vector, 
$\mathbf{E}(\mathbf{x},t)$,  
up to 18 MHz, with selective Nyqvist frequency bandwidths down to 5 
kHz, and one component of the magnetic field 
vector, $\mathbf{B}(\mathbf{x},t)$, from a few Hz up to 100 kHz. 
As a transient detector, 
ELVIS is capable of detecting pulses with a minimum pulse width of 5 ns.
The instrument comprises three orthogonal electric dipole antennas, 
one magnetic search 
coil antenna and a four-channel digital sampling system, utilising 
flexible digital down conversion and filtering together with 
state-of-the-art onboard digital signal processing.
\end{abstract}

\thispagestyle{empty}
\begin{spacing}{1}

\section*{The ELVIS Instrument}
\begin{figure}
\includegraphics[width=\columnwidth]{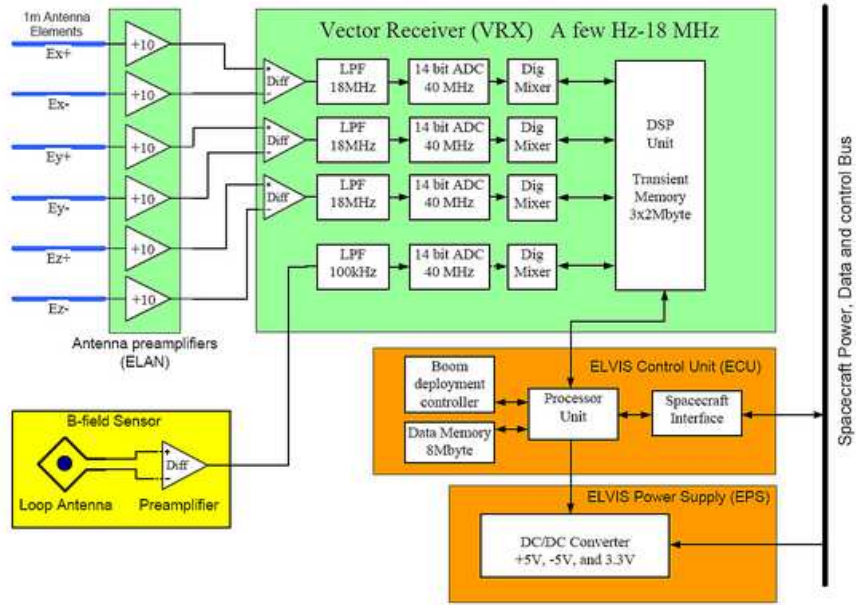}
\caption{ELVIS block diagram}
\label{fig:ELVIS}
\end{figure}
Vector measurements are critical for an unambiguous characterisation of
electromagnetic fields and their polarisation \cite{carozzi2000}, including wave direction finding.
Therefore, ELVIS comprises three, 5 m long, electric dipole antennas, 
mounted on the satellite body in an as close as 
possible orthogonal configuration. 
Each dipole is connected to a digital receiver, which 
allows for detailed investigation of 
electric field vector fluctuations up to 18 MHz.
As a transient detector, 
ELVIS has a time resolution of 12.5 ns and a detection limit of 5 ns.
For high-sensitivity measurements below 100 kHz, 
a magnetic search-coil antenna, 
measuring one component of the magnetic field vector,
is connected to a fourth, identical, digital receiver. 
Fig. \ref{fig:ELVIS} illustrates 
schematically the functionality of the ELVIS instrument. 

One electric dipole antenna is created by 
connecting two active, 2.5 m long, 
mono\-pole antennas 
to a differential amplifier. The 
digital receiver then  contains
three stages: 
The first stage is an analogue anti-aliasing filter. Lowpass, 
19$^{\rm{th}}$ order, Chebyshev filters have been chosen here. The second stage 
comprises the 
analogue-to-digital conversion (ADC). Synchronised, high speed 14 bit ADCs 
with
a sampling
frequency of 40 MHz are used. Operated as a transient detector
the anti-aliasing filters are bypassed and the ADC sampling frequency is doubled
to 80 MHz. The ADCs are limited to 500 MHz input frequency, but it is
the ELVIS low-noise antenna pre-amplifiers (ELNA) that limits
the detectability. This limit is 200 MHz,
corresponding to 5 ns pulses. 
The third stage of the receiver is a four-channel digital downconverter (DDC). 
The DDC is the core of ELVIS, as it digitally performs all traditional, 
analogue,
receiver functions.
It is a, so-called, \emph{hard-wired} digital signal processor (DSP), 
which is optimised to perform
digital frequency mixing, filtering and sampling-rate reduction. 
Reduction factors between 1 (full bandwidth) and 4096 can be used.  
This is reflected in the dynamic range of ELVIS, which is 81--84 dB (14 bits)
at full
bandwidth and 117--120 dB (20 bits) at a reduction of 4096, 
corresponding to 5 kHz bandwidth
at 40 MHz sampling frequency. Nominally, ELVIS has a sensitivity of -135 dBm/Hz. 
An additional DSP for data post-processing and a 3$\times$2 MB transient memory makes
up the full vector receiver (VRX), which uses one printed circuit board
(PCB). A second PCB houses the ELVIS controller unit (ECU), 
and power
supply (EPS). The ECU is the only control and data interface to the spacecraft. 
For the Chandrayaan-1 mission, the ELVIS  telemetry requirement is
128 kbps.
Besides the
processor unit the ECU contains a dedicated boom deployment controller and 
8 MB of memory for 
processed data. The
EPS is a DC/DC converter that supplies +5V, -5V and 3.3V to ELVIS and  
it is the only power interface to the spacecraft. Normally, ELVIS requires 4W 
of power but during transient detection this is increased to 7W. A peak
value of 10W is expected during short periods. 

\begin{figure}
\includegraphics[width=\columnwidth]{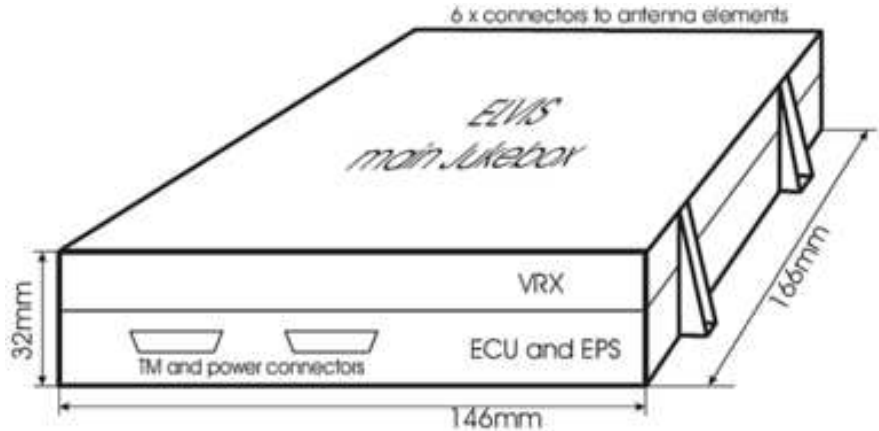}
\caption{ELVIS main electronics box (JUKEBOX)}
\label{fig:JUKEBOX}
\end{figure}

Mechanically, the VRX and the
combined ECU/EPS
PCBs are housed in the ELVIS main electronics box (JUKEBOX), as
shown in Fig. \ref{fig:JUKEBOX}. The JUKEBOX is made of
two NCN-milled aluminium frames 
and measures 166$\times$146$\times$32 mm$^3$.
The base plates are 0.5 mm thick with 4 mm thick stiffening
covering 20 \% of the box surface area. The mass of the complete
electronics box, including PCBs, screws and connectors, is 630 gram.
The six electric antenna elements (EANT) are made of a metal foil that is 
stored on a reel and forms 2.5 m long tubular booms when deployed. The
mass of one unit is 150 grams, with the antenna tube itself weighing
less than 50 grams.  
To reduce the magnetic contamination from instruments and systems onboard 
the spacecraft, the search coil sensor will be installed on the top 
of a deployable rigid boom, at a minimum distance of 500 mm from the 
spacecraft. For deployment, the boom has a bellow hinge that is filled with
a low melting-point metal. A heater is used to melt the metal and operate the
hinge. The mass of the search-coil sensor, including the boom unit, is 150
grams. The total mass of the ELVIS instrument is 1630 grams, excluding
cable harness.

\begin{figure}
\includegraphics[width=\columnwidth]{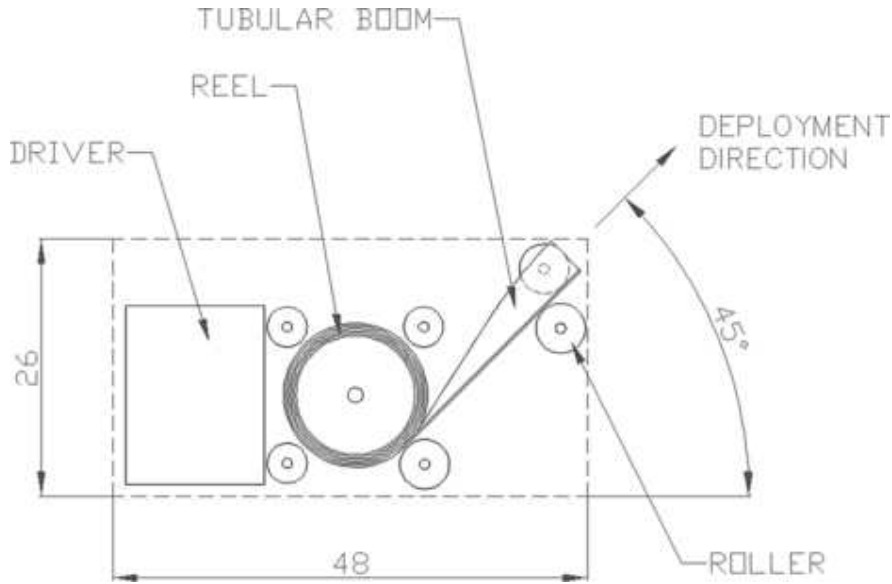}
\caption{Electric Antenna deployment unit (NanoSpace-1 design by J. Grygorczuk,
Space Research Centre, Warsaw, Poland). 
Scale in mm.}
\label{fig:EANT}
\end{figure}

At a cost of performance, using shorter antennas the ELVIS mass can be 
reduced quite substantially. 
On the Swedish NanoSpace-1 (NS1) nanosatellite, 
1 m antenna elements, weighing only 25 grams each, will be used, see
Fig. \ref{fig:EANT}. For
ELVIS onboard Chandrayaan-1 that would
save 750 grams. Another 100 grams can be saved by using Mg rather than
Al as material for the JUKEBOX. Together, these savings gives a 
total mass of 730 grams. To further reduce mass significantly, one has to
replace the PCB electronics. One alternative that we will use on NS1
is a so-called
\emph{multi chip module} (MCM), where naked electronics chips are 
embedded directly in a 
single Silicon wafer. In this case a 7$\times$7 cm$^2$ wafer with a mass
of 25 grams is used to accommodate ELVIS and three other instruments.

\section*{Scientific Objectives}

A thorough knowledge of the electromagnetic environment of the Moon is a 
necessity before low-frequency radio observations from the Moon
can be realised, which is our long-term goal.
The scientific objective of 
ELVIS is therefore to explore the electromagnetic environment 
of the 
Moon. To this end, investigations of the electromagnetic field 
properties, their sources and origins will be carried out.

Radio measurements are very efficient and versatile and can be used to 
study the Moon \emph{itself} and its interaction with the surrounding space 
plasma and solar wind. 
For studies of remote radio sources, including radio astronomy,
using the Moon 
as a \emph{shield}, the far side of the Moon is the most radio quiet site 
within reach in our solar system. That is why a low frequency radio 
telescope on the far side of the Moon has been a dream for many decades.  
Using instead the moon as a \emph{target}, 
there are also more exotic emissions of cosmic origin that remains 
to be studied. 
Of particular interest are those caused by ultra-high energy 
neutrinos 
impinging upon
the lunar surface, which
give rise to transient
radio pulses through the Askaryan effect [1]. Such pulses
can be detected by ELVIS.

\subsection*{Lunar Space Plasma}
A whole range of unexplored plasma phenomena are foreseen in the 
neighbourhood of the Moon. 
Most of our knowledge of the lunar plasma comes from 
a lunar wake passage at  6.5 R$_{\rm{L}}$ by
the WIND spacecraft on Dec. 27, 1994
\cite{ogilvie96, bosqued96, owen96, kellogg96}. 
The WIND 
observations revealed a lunar wake density cavity with an electron density
below 0.2 cm$^{-3}$,
where there existed a heated electron 
population of up 
to a few 100 eV, as well as a cross wake ion flow. 

The cross wake current
has to close somewhere near the Moon. 
One possibility 
is that electric currents are carried by a dense 
photo electron layer \cite{vyshlov76, vyshlov78, bauer96}  
near the dayside lunar surface. 
A most interesting aspect 
of the solar wind -- Moon interactions observed by WIND 
was the detection of a 
multitude 
of plasma and radio emissions in the 
lunar wake as well as in the wake lobes 
\cite{kellogg96}. 
The radio emissions observed inside the wake matched the electron 
plasma frequency 
observed at its edges. 
(The angular electron plasma frequency in SI units is given by
$\omega_{\rm{pe}}=\sqrt{q_{\rm{e}}^2 n/\epsilon_0 m_{\rm{e}}}$, 
where $n$ 
is the number density in m$^3$.)
It is believed that these radio emissions 
originate in the turbulent plasma near the edges of the wake.

When the Moon passes the night side of the Earth, measurements can be 
made of the 
properties of the geomagnetic tail, preferably in coordination with other 
magnetospheric 
satellites. 
An ELVIS instrument in orbit around the Moon can detect large-scale plasma 
cavities
when these, 
as a result of 
geomagnetic storms, 
move outward in the magnetospheric tail 
and pass the Moon. What effect the Moon plasma 
itself has on the geomagnetic tail and the near Earth magnetospheric processes 
is not known. If a well developed lunar ionospheric plasma exists, 
significant effects on 
the geomagnetic tail is expected because this plasma will act both as a 
mass load and as 
a diversion of electrical currents in the tail.

There appears to exist mini-magnetospheres on the Moon,
as shown recently \cite{futaana2003}. 
It is believed that these are
caused by local magnetic anomalies in the lunar crust.
This
gives rise to the possibility of small scale
shocks, 
which could be potentially active sites for wave -- particle interaction 
processes and radio emissions \cite{kuncic2004}. 
Such radio emissions could be detectable by ELVIS, thus verifying 
the presence of the mini-magnetospheres. 

\subsection*{LOFAR/LOIS and Radio Observatories on the Moon}
A gigantic network of sensors for astronomy, space and environmental studies 
is being built in Europe. Sweden has been part of this cooperation for several 
years by contributing with sensor, IT and radio research, as well as a test 
compound in southern Sweden. 
The heart of this trans-European sensor network is LOFAR 
[www.lofar.org], 
currently being built in the Netherlands. Data streams at up to Terabits 
per second from tens of thousands of small, inexpensive sensors, placed 
in an area about 300 km across, and connected by fibre-optic networks, 
are processed in powerful supercomputers. 
This results in the world's most sensitive radio telescope in the hitherto 
unexplored lowest frequency band 10-250 MHz. In addition to space research, 
the sensor network will be used for geophysics, agricultural and environmental 
research and as a test-bed for future wireless and optical fibre networks. 

The LOIS [www.lois-space.net] subproject was born in 2000.
The objective is to build a space-, telecom- and IT-research 
supplement to LOFAR in southern Sweden. The LOFAR/LOIS combination will be 
a next-generation network-based radar for solar system studies that will 
provide opportunities to make research at the cutting edge of radio-based 
space science.

A low-frequency lunar radio telescope would be realised using a modern phased 
array concept, pioneered in Europe by LOFAR/LOIS, consisting of about 100 
light-weight tripole antennas connected by a wide area sensor network.

The ELVIS instrument can demonstrate the performance 
of one such antenna and act as a precursor for a lunar radio 
observatory. In addition, coordinated measurements with ELVIS and LOFAR/LOIS 
would be able to produce scientific results of high significant value.

\subsection*{Ultra-high Energy Neutrino Detection}
As described in a companion paper \cite{Stal05}, ELVIS could potentially
detect radio emissions caused by ultra-high energy neutrinos hitting the Moon.
Neutrinos are difficult to detect since they interact with matter only 
through weak processes with low cross-section. To access the decreasing 
flux of ultra-high energy (UHE) cosmic neutrinos with an energy far 
above 1 PeV, target volumes well in excess of 1 km$^3$ are required. 
Askaryan proposed early \cite{Askaryan62, Askaryan65} that, for 
these energies, it might be advantageous to look for the coherent, radio 
frequency, Vavilov-\v{C}erenkov emission. The coherent emission is 
generated by a negative charge excess in the showers of secondary 
particles that is produced when the neutrino interacts with a dense 
material \cite{Askaryan62, SLAC}. 

Such interactions should take place in the lunar regolith, and searches 
for Askaryan pulses have been performed using radio telescopes 
on Earth \cite{GLUE}. We believe that such pulses could be detectable 
by the ELVIS instrument, considering especially the $70$ dB gain the 
closer distance to the moon provides compared to Earth-based observations. 
Simulations on the expected detection efficiency for UHE neutrino events 
have been performed \cite{Stal05}, showing a neutrino energy threshold 
of $5\times10^{19}$ eV and a model dependent expected event rate 
of $2.2$ detectable events per year of observation time.

\section*{Measurement Principles}
To describe electromagnetic waves, an analytic representation in terms of 
complex valued three-dimensional electric and magnetic vector fields, 
$\mathbf{E}(\mathbf{x},t)$ and $\mathbf{B}(\mathbf{x},t)$, respectively 
is convenient. The analytic representation 
has several advantages, and is the natural choice for physicists as well
as radio engineers.
In ELVIS, the analytic representation is obtained in the DDC, which outputs
complex time-series, in terms of
16 bit in-phase and quadrature-phase ({\it{I}},{\it{Q}}), components.

In general, by polarisation we mean the 2$^{\rm{nd}}$ order statistics of any
combination \cite{carozzi2005} of the complex
field vectors $\mathbf{E}$ and $\mathbf{B}$. 
ELVIS measures only
$\mathbf{E}$, and therefore 
the coherency (or spectral) density matrix can be used:
\begin{align}
\mathbf{E}\mathbf{E}^{\dagger}
=\left(
\begin{array}{ccc}
E_x E_x^{\ast} & E_x E_y^{\ast} &E_x E_z^{\ast}\\
E_y E_x^{\ast} & E_y E_y^{\ast} &E_y E_z^{\ast}\\
E_z E_x^{\ast} & E_z E_y^{\ast} &E_z E_z^{\ast}
\end{array}
\right)
\label{eq:S}
\end{align}
This matrix completely describes $\mathbf{E}$-field polarisation 
in 3D and is a generalisation \cite{carozzi2000} of
Stokes 2D description. For instance, the intensity, 
$\mathbf{E}\cdot\mathbf{E}^{\ast}$, is the trace and
the imaginary part of the off-diagonal
elements yields a
vector,  
$\rm{i}\mathbf{E}\times\mathbf{E}^{\ast}$,  
that in vacuum is parallel to the direction of wave propagation.
Furthermore, a measure of the degree of circular  polarisation is given by
the ratio
$|\rm{i}\mathbf{E}\times\mathbf{E}^{\ast}|/\mathbf{E}\cdot\mathbf{E}^{\ast}|$.
In 3D there are altogether five independent instantaneous polarisation 
parameters. By taking averages of the matrix in (\ref{eq:S}) the full set of
nine polarisation parameters can be used. In 2D, these parameters reduces to 
the four well-known 
Stokes parameters.

Sampling at 40 MHz, ELVIS will produce 2 Gbps of raw data. 
Clearly, this huge
data rate must be reduced. A major reduction, down to 1 Mbps at 5 kHz bandwidth
can be performed by the DDC but this is not enough. Post processing of data
in the DSP is therefore necessary. The post processing is a trade-off between
time, frequency and polarisation coverage. On ELVIS, the field entropy is
calculated regularly and passed to housekeeping (HK) data. Entropy is a
measure of information that can be used to turn on and off different modes
according to pre-defined rules. It
can also be used as a signal for other instruments onboard the spacecraft.

ELVIS operates in four basic modes: Survey Mode (SM),  Normal Mode (NM)
Burst Mode (BM), and Transient Mode (TM). The time allocation for these 
modes are 60\%,  15\%,  15\%, and 10\%, respectively. In SM, ELVIS performs
narrowband frequency sweeps and in NM it
operates as a baseband {\it{I}}/{\it{Q}} receiver. 
The BM and TM are similar, capturing
the full bandwidth waveform, but in TM the
sampling frequency is
doubled to
80 MHz and the Nyquist filters are bypassed. Only those events identified as 
transients are stored.
\section*{Discussion}
The design philosophy of ELVIS is to eliminate all unnecessary
analogue electronics. The receiver functionality is instead performed digitally.
This minimises signal distortion, 
reduces mass and makes the instrument more robust compared to analogue designs. 
However,
the major advantage of ELVIS
lies in its
high flexibility, which makes it possible to
address the broad spectrum of scientific objectives outlined here,
including lunar space plasma studies, remote sensing, and ultra-high
energy neutrino detection.

ELVIS builds on the heritage of proven space 
science instruments
constructed at the Swedish Institute of Space Physics (IRF) in Uppsala,
which currently has instruments in operation  onboard six satellites. These are
the four ESA Cluster satellites, the NASA Cassini
spacecraft, and the ESA Rosetta spacecraft.
ELVIS itself has no flight-record yet but will have very soon.
A 
one-channel version will
be launched  before the end of 2005 onboard the Russian Compass-2 satellite, 
aimed at investigating
precursors to earthquakes. A similar instrument but with three channels,
will be installed by the end of 2006 on the 
International Space Station 
(ISS),
for monitoring the ISS space environment.
Eight digital vector receivers, identical to the ISS receiver, are  
since 2004 in operation
at the LOIS Test Station in V\"axj\"o, Sweden. An equivalent receiver at
IRF in Uppsala, has worked flawlessly in an outdoor setting 
for more  than three years.
A miniaturised ELVIS instrument MCM is also being developed for the Swedish
NanoSpace-1 satellite to be launched in 2007--2008. 
To be selected for future grand missions, our strategy is to gain 
flight-experience onboard low-cost
microsatellites. One satellite per year is foreseen 
in the next few years.

\end{spacing}
\bibliographystyle{prsty}
\bibliography{lurefs}
\end{document}